\pdfoutput=1

\documentclass[preprint]{aastex}

\usepackage{amsmath}
\usepackage{graphicx}

\newcommand{\F}{\mathcal{F}}
\newcommand{\Fstar}{\F^{\star}}
\newcommand{\f}{f}
\newcommand{\fdot}{\dot{f}}
\newcommand{\fdotdot}{\ddot{f}}
\def\LIGOdcc{LIGO-P1000028-v7}

\begin{document}

\title{First search for gravitational waves from the youngest known neutron star}
\shorttitle{Search for GWs from the youngest known NS}

% LSC Author List, October 2009 (LIGO-T0900658-v1)
% Based on "authors_list_october2009_ligo_byAuthor.tex"
% Changes:
%   - Author affiliation marks changed to revtex macros
%   - Added principal affiliation to LSC

\author{
J.~Abadie\altaffilmark{1}, 
B.~P.~Abbott\altaffilmark{1}, 
R.~Abbott\altaffilmark{1}, 
M,~Abernathy\altaffilmark{2}, 
C.~Adams\altaffilmark{3}, 
R.~Adhikari\altaffilmark{1}, 
P.~Ajith\altaffilmark{1}, 
B.~Allen\altaffilmark{4,5}, 
G.~Allen\altaffilmark{6}, 
E.~Amador~Ceron\altaffilmark{5}
% ...
}

\affil{The LIGO Scientific Collaboration, http://www.ligo.org}

\altaffiltext{1}{LIGO - California Institute of Technology, Pasadena, CA  91125, USA }
\altaffiltext{2}{University of Glasgow, Glasgow, G12 8QQ, United Kingdom }
\altaffiltext{3}{LIGO - Livingston Observatory, Livingston, LA  70754, USA }
\altaffiltext{4}{Albert-Einstein-Institut, Max-Planck-Institut f\"ur Gravitationsphysik, D-30167 Hannover, Germany}
\altaffiltext{5}{University of Wisconsin--Milwaukee, Milwaukee, WI  53201, USA }
\altaffiltext{6}{Stanford University, Stanford, CA  94305, USA }
% ...

\shortauthors{The LIGO Scientific Collaboration}

\begin{abstract}
  We present a search for periodic gravitational waves from the neutron star in the supernova
  remnant Cassiopeia A.  The search coherently analyzes data in a 12-day interval taken from the
  fifth science run of the Laser Interferometer Gravitational-Wave Observatory.  It searches gravitational wave frequencies from 100 to 300~Hz, and covers a
  wide range of first and second frequency derivatives appropriate for the age of the remnant and
  for different spin-down mechanisms.  No gravitational wave signal was detected.
  Within the range of search frequencies, we set 95\% confidence upper limits of
  $0.7$--$1.2 \times 10^{-24}$ on the intrinsic gravitational wave strain,
  $0.4$--$4 \times 10^{-4}$ on the equatorial ellipticity of the neutron star, and
  $0.005$--$0.14$ on the amplitude of $r$-mode oscillations of the neutron star.  
  These direct upper limits beat indirect limits derived from
  energy conservation and enter the range of theoretical predictions involving crystalline exotic matter or runaway $r$-modes.
This paper is also the first gravitational-wave search to present upper
limits on the $r$-mode amplitude.
\end{abstract}


\section{Introduction}\label{sec:intro}

Using data from the Laser Interferometer Gravitational-Wave Observatory
\citep[LIGO;][]{LSC-det-2009}, the LIGO Scientific Collaboration (LSC) and Virgo Collaboration have published searches for
periodic gravitational waves from three astrophysically distinct types of rapidly rotating neutron star.
Searches have targeted non-accreting pulsars
\citep{LSC-CW-known-S1,LSC-CW-known-S2,LSC-CW-known-S3S4,LSC-CW-known-S5}, most notably the Crab
pulsar \citep{LSC-CW-Crab-S5,LSC-CW-Crab-S5-err,LSC-CW-known-S5}, using data from LIGO's first five science runs
(designated S1--S5).  Two searches have targeted the accreting neutron star in the low-mass x-ray binary Scorpius X-1,
using data from S2 \citep{LSC-CW-Fstat-S2} and S4 \citep{LSC-stochastic-map-S4}.  Other searches
have been broadband all-sky surveys for as-yet undiscovered neutron stars, using data from S2
\citep{LSC-CW-Hough-S2,LSC-CW-Fstat-S2}, S4 \citep{LSC-CW-PSH-S4,LSC-CW-EatH-S4}, and S5
\citep{LSC-CW-PF-S5,LSC-CW-EatH-S5}.

In this paper, we present the first directed search for periodic gravitational waves from a known,
isolated, non-pulsing neutron star.  The search targets the central compact object (CCO) in the
supernova remnant Cassiopeia A (Cas~A).
The remnant
is estimated to be $3.4_{-0.1}^{+0.3}$~kpc distant
\citep{Reed-etal-1995}, and to have been born in the year $1681 \pm 19$ \citep{Fesen-etal-age-2006}.
It is the second-youngest known supernova remnant in the Galaxy, and the youngest with a confirmed
CCO \citep{Reynolds-etal-2008,DeLuca-2008}. 
The remnant and CCO have been extensively studied through electromagnetic
observations, which we summarize in Section~\ref{sec:intro:em}.

There is compelling evidence that the Cas~A CCO is a neutron star
\citep{Pavlov-etal-2000,Chakrabarty-etal-2001,Gotthelf-Halpern-2008,Pavlov-Luna-2009,Ho-Heinke-2009}.
We argue, in Section~\ref{sec:intro:gw}, that its age and youth make it an
interesting target for a search for periodic gravitational waves.
Its youth means that it has not been covered by all-sky surveys for periodic
gravitational waves, which focus on spin-down timescales much longer than the
age of the remnant
\citep{LSC-CW-Hough-S2, LSC-CW-Fstat-S2, LSC-CW-PSH-S4, LSC-CW-EatH-S4,
LSC-CW-PF-S5, LSC-CW-EatH-S5}.
We describe the first search for gravitational waves from Cas~A in
Section~\ref{sec:search}, present the results of the search in
Section~\ref{sec:results}, and discuss them in Section~\ref{sec:discuss}.
(We shall often abbreviate ``the Cas~A central compact object'' to ``Cas~A.'')

Since the rotation frequency, and hence the gravitational wave frequency, of
Cas~A is unknown, we search for periodic gravitational waves with frequencies between 100~and 300~Hz.
At these frequencies, where the strain noise of the LIGO detectors is lowest, 
the search is designed to beat indirect upper limits on
gravitational radiation based on energy conservation \citep[see][and Sections~\ref{sec:intro:gw} and~\ref{sec:search:param}]{CasA-methods-2008}.
The search found no credible signal (see Section~\ref{sec:results}).
In the absence of a detection, we present 95\% confidence upper limits on a gravitational wave signal from Cas~A, assuming its frequency is within the searched band. Upper limits are given for 
the intrinsic gravitational wave strain~$h_0$, the equatorial ellipticity~$\epsilon$,
and the $r$-mode amplitude~$\alpha$.

Within the searched frequency band, the upper limits presented in this paper beat the
indirect upper limits, as expected.  Cas~A is now one of
only a handful of neutron stars \citep[see][]{LSC-CW-known-S5} where the most sensitive upper limits
on gravitational radiation have been obtained using gravitational wave detectors such as LIGO.
This paper is also the first gravitational-wave search to present upper
limits on the $r$-mode amplitude.
The best upper limits on $\epsilon$ (a few times $10^{-5}$) and $\alpha$ (a few
times $10^{-3}$) are within the range of some theoretical predictions (see
Section~\ref{sec:discuss}).

\subsection{Electromagnetic Observations}\label{sec:intro:em}

The Cas~A CCO was first discovered as an x-ray point source in first-light images taken by the
Chandra X-ray Observatory \citep{Tananbaum-1999}.  It was subsequently identified in other satellite
data dating back to the year 1979; the x-ray flux appears to have been constant since then
\citep{Pavlov-etal-2000}.
Optical and near-infrared searches have not found the CCO, and have all but ruled out the presence of
an accretion disk from fallback, and a binary companion
\citep{Ryan-etal-2001,Kaplan-etal-2001,Fesen-etal-limits-2006,Wang-etal-2007}.
The absence of the latter is
puzzling, since the light-echo spectrum of the supernova, of type IIb,
implies that the progenitor was stripped of hydrogen by a companion
\citep{Young-etal-2006, Krause-etal-supernova-2008}.
One possibility is that the progenitor was the product of binary companions
merging during a common envelope phase \citep{Krause-etal-supernova-2008}.

The CCO appears to be a neutron star with a low surface magnetic
field (an anti-magnetar).
Blackbody fits to the x-ray spectrum by \citet{Pavlov-etal-2000} and \citet{Chakrabarty-etal-2001}
implied a temperature too high, and an
emitting area too small, to be consistent with emission from the whole surface of a cooling neutron
star, or from the inner region of an accretion disk around a black hole.
\citet{Pavlov-etal-2000} proposed a model of a strongly magnetized neutron star
(a magnetar) with hot polar caps, which would, however, lead to x-ray pulsations
which have not been observed (see below).
Light echoes from the explosion have been interpreted as signs of a flare,
reminiscent of a soft gamma repeater, occuring in the year 1953
\citep{Krause-etal-echoes-2005}.
Such a flare, if single, could also have been due to a one-time phase
transition \citep{Gotthelf-Halpern-2008}.
More recently, however, this interpretation of the echoes, and thus the
evidence for the flare, has been discounted \citep{Kim-etal-2008,
Dwek-Arendt-2008}.
Recently, \citet{Ho-Heinke-2009} combined previous x-ray spectra
\citep{Hwang-etal-2004, Pavlov-Luna-2009}, carefully adjusted for instrumental
effects, and fitted them to various light-element atmosphere models.
Their best fit was for nearly isotropic emission from the entire surface of a
neutron star with a carbon atmosphere, low magnetic field, and mass and radius within normal ranges.

The spin period of the neutron star is unknown.
\citet{McLaughlin-etal-2001} searched for radio pulses (including possible
binary orbital periods as short as a few hours) and found none at periods as
short as 1--10~ms, depending on dispersion measure, making the CCO much more
radio quiet than any known radio pulsar under $10^4$ years old.  Searches for
X-ray pulsations at periods
as short as 2~ms
\citep{Chakrabarty-etal-2001,Murray-etal-2002,Mereghetti-etal-2002,Pavlov-Luna-2009} have produced
at best a marginal candidate for a period at 12~ms, which has not been confirmed.
No pulsar wind
nebula has been detected \citep{Hwang-etal-2004,Pavlov-Luna-2009}.

\subsection{Motivation for a Gravitational Wave Search}\label{sec:intro:gw}

If the CCO in Cas~A is an anti-magnetar, it may be spinning fast enough to emit
periodic gravitational waves above 100 Hz, where LIGO is most
sensitive.

To date, seven supernova remnant CCOs are known: three have observed spin
periods, and a fourth has an observed periodicity that may be due to binary
orbital motion \citep{DeLuca-2008,Gotthelf-Halpern-2009}.
The fastest-spinning CCOs have
rotation periods of $\sim 100$~ms, which are much slower than the longest rotation periods covered
by this search (20~ms for gravitational waves from a non-axisymmetric distortion, and 13~ms for
$r$-modes; see Section~\ref{sec:results}).
Only one of the CCOs has a measurable spin-down \citep{HalpernGotthelf2010}
and the others have tight upper limits.
This indicates that the CCOs have magnetic fields much less than those of typical radio pulsars,
and rotation periods that have not changed significantly since birth.
If Cas~A spins as slowly and constantly as these objects, it is not detectable
by LIGO.
It is difficult, however, to definitively extrapolate the general properties of
CCOs from a sample of three or four.

Young neutron stars, such as Cas~A, may be the most likely to retain
non-axisymmetries from the violent circumstances of their births.
For example, the formation of the crust during an epoch of perturbations such
as $r$-modes \citep{Lindblom-etal-2000, Wu-etal-2001} could lead to an
irregular shape.
Gravitational radiation may also be generated by the continuing non-axisymmetry
of the $r$-modes themselves \citep{Owen-etal-rmode-1998}, which may last for up
to thousands of years \citep{ArrasEtAl2003} depending on the composition and
viscosity of the star.

Indirect upper limits on gravitational wave emission from Cas~A can be estimated using a method similar to
the spin-down limit for known pulsars \citep{CasA-methods-2008, rpol}.  If the star was born spinning at least $\sim20\%$
more rapidly than it is now, and the spin-down evolution has been dominated by the emission
of gravitational waves, the unknown spin frequency and frequency derivative can be eliminated in
favor of the known age to place a rough upper limit on the gravitational wave emission.  For Cas~A,
the indirect limit on the intrinsic gravitational wave strain $h_0$ (see Section~\ref{sec:search:Fstat}) is
\begin{equation}
\label{eqn:ul-h0}
h_0 \lesssim 1.2 \times 10^{-24} D_{3.4}^{-1} \tau_{300}^{-1/2} I_{45}^{1/2}
\,,
\end{equation}
where $D_{3.4}$ is the distance to Cas~A in units of 3.4~kpc, $\tau_{300}$ is
its age in units of 300~yr, and $I_{45}$ is its principal moment of inertia in
units of $10^{45}\textrm{ g cm}^2$.
The indirect limit on $h_0$ is independent of frequency.

The choice of a fiducial age of 300~yr for Cas~A, at the young end of the range estimated
by \citet{Fesen-etal-age-2006}, is conservative in that it gives a larger
search parameter space (see Section~\ref{sec:search:param}).
It also raises the indirect limit by $\sim 10\%$, a small effect compared to the
uncertainties in the distance (of order 10\%), and
the principal moment of inertia, which may be up to 3 times higher than its fiducial
value \citep[see][]{LSC-CW-known-S3S4}.
The uncertainties in the direct upper limits presented in
Section~\ref{sec:results} are on the order of 10--15\%, due to
uncertainties in the calibration of the LIGO detectors (of order 10\%; see Section~\ref{sec:search:data}),
and systematic uncertainties in the search pipeline (of order 5\%; see Section~\ref{sec:search:Fstat}).
Equation~\eqref{eqn:ul-h0} assumes
gravitational waves from a mass quadrupole; the equivalent limit for $r$-modes
is higher by tens of percent \citep{rpol}.

The indirect limit on $h_0$ may be converted into an indirect limit on the
equatorial ellipticity
\begin{equation}
\epsilon \lesssim 3.9 \times 10^{-4} \tau_{300}^{-1/2} I_{45}^{-1/2}
f_{100}^{-2} \,,
\label{eqn:ul-eps}
\end{equation}
where $f_{100}$ is the gravitational wave frequency in units of 100~Hz.
Due to the uncertainties in neutron star parameters mentioned above, this limit
on $\epsilon$ is overall uncertain by roughly a factor 2.
The indirect limit on $h_0$ also implies an indirect limit on gravitational
waves from $r$-modes \citep{rpol}, in terms of their amplitude
\begin{equation}
\alpha \lesssim 0.14 \tau_{300}^{-1/2} f_{100}^{-3} \,,
\label{eqn:ul-alpha}
\end{equation}
where the uncertainty due to the moment of inertia and other properties of the
star, while more complicated, is roughly a factor of~2--3.
If $\alpha$ varies over the observation time, this limit applies to the rms
value of $\alpha$ over time.
See \citet{rpol} for precise definitions of $\alpha$ and $\epsilon$,
translations to other quantities used in the literature (including $h_0$), and
more discussion of uncertainties.

\citet{CasA-methods-2008} showed that a search for Cas~A of 12 days of
LIGO S5 data in the band 100--300~Hz is feasible and can expect to beat the
indirect limits.
The Cas~A indirect limits on $h_0$ and $\epsilon$ are comparable to the
spin-down limits for the Crab pulsar, which have been beaten by searches of
LIGO S5 data \citep{LSC-CW-Crab-S5,LSC-CW-known-S5}.
The indirect limits on $\alpha$ are lower and therefore more interesting than
for the Crab if Cas~A is emitting at 100--300~Hz.
The indirect limits are also comparable to the best upper limits achieved by
all-sky searches for periodic gravitational waves \citep{LSC-CW-PF-S5,
LSC-CW-EatH-S5}.
The all-sky searches covered longer spin-down timescales and thus were not
sensitive to Cas~A if it is emitting gravitational waves near the indirect
limit.

\section{Search Pipeline}\label{sec:search}

This section describes the gravitational wave search for Cas~A. The search pipeline consists of:
the selection of data for the search (Section~\ref{sec:search:data}), the analysis method
(Section~\ref{sec:search:Fstat}), the search parameter space (Section~\ref{sec:search:param}),
the template bank used to compute the search (Section~\ref{sec:search:templ}), post-processing of the results
(Section~\ref{sec:search:post}), examination of the significance of the largest value of the
detection statistic returned by the search (Section~\ref{sec:search:2Fstar}), and, in the event of no detection, calculation of the upper limits
(Section~\ref{sec:search:ul}).

\subsection{Data Selection from the LIGO S5 Run}\label{sec:search:data}

LIGO is a network of three interferometric detectors: a 4-km arm-length detector in Livingston,
Louisiana (L1) and two detectors of 4-km (H1) and 2-km (H2) arm lengths co-located in Hanford,
Washington.

The S5 science run \citep{LSC-det-2009} is LIGO's fifth and most recently completed science run.  It commenced at 2005
November 4, 16:00 UTC at Hanford, and at 2005 November 14, 16:00 UTC at Livingston; it ended at 2007
October 1, 00:00 UTC.  The S5 run collected over one year of science data coincident among all
three detectors, with an overall triple-coincidence duty cycle of 54\%.  Interruptions caused by
environmental disturbances, as well as scheduled breaks for maintenance and commissioning of
equipment, accounted for the downtime.  During S5, the detectors were operating at very near their
design sensitivities.  The strain noise of the two 4-km detectors was on average less than
$3 \times 10^{-23} \textrm{ Hz}^{-1/2}$ at their most sensitive frequencies (around 140 Hz) and less
than $5 \times 10^{-23} \textrm{ Hz}^{-1/2}$ over 100--300 Hz, and generally improved (as did the
duty cycle) over the course of the run.

This search uses science data from only the L1 and H1 detectors.
Data from the H2 detector is less
sensitive, but carries the same computational cost to search.  A small percentage of the acquired
science data is excluded by data quality controls,
which identify times when the data is known to be unsuitable for analysis.
This includes, e.g., data taken when the output
photodiodes of an interferometer were saturated, when the calibration of the data was ill-defined,
when high winds were measured at the Hanford observatory, and 30 seconds before an interferometer lost
lock.  The remaining science data is calibrated \citep{LSC-CW-known-S5,LSC-det-2009,LSC-calib-2010} to produce a
(discontinuous) time series of gravitational wave strain, $h(t)$.
The time series is then broken into 30-minute segments.
Because not every continuous section of $h(t)$ is an integer multiple of 30 minutes in length,
some science data is discarded in this process.
Finally, each 30-minute segment is high-pass filtered above 40~Hz 
and Fourier transformed\footnote{
Prior to the Fourier transform, each segment is multiplied by a nearly-square Tukey window
to mitigate transients at the start and end of the segment.
The loss in signal power due to the windowing is on the order of 0.1\%.
} to
form Short Fourier Transforms (SFTs) of $h(t)$.  The SFTs are the input data to the search pipeline.

The maximum uncertainties in the calibration of $h(t)$ are
10.4\% in amplitude and $4^{\circ}.5$ in phase for H1, and
14.4\% in amplitude and $4^{\circ}.2$ in phase for L1;
all uncertainties are constant in time to within 1\% \citep{LSC-calib-2010}.
The uncertainties in the calibration amplitude contribute to the overall systematic
uncertainty in the upper limits presented in Section~\ref{sec:results}.
The analysis method used in this search (see Section~\ref{sec:search:Fstat}) is sensitive
only to the relative phase uncertainty between detectors.
Even for a worst-case relative phase uncertainty of $\sim 10^{\circ}$, the resulting 
difference in the signal frequency between detectors would still be much less than the mismatch 
allowed for by the search template bank (see Section~\ref{sec:search:templ}).
Thus, the phase uncertainties do not affect the results of this search.

The search is restricted to a data set spanning a maximum of 12 days \citep{CasA-methods-2008}.
The computational cost of a coherent search for Cas~A scales with the 7th power
of the timespan between the first and last timestamp of the analysed SFTs,
while the sensitivity scales only with the square root of the observation time.
Increasing the timespan from 12 days would therefore rapidly increase the
computational cost, for a negligible improvement in sensitivity.
To select the 12-day data set, we compute a figure of merit
for each possible data set spanning 12 days, chosen from the available S5 SFTs.
The figure of merit, which
is proportional to the estimated power signal-to-noise ratio, is given by
$\sum_{k,f} [S_h(f)]^{-1}$, where $S_h$ is the strain noise power spectral
density at frequency $f$ in the SFT numbered $k$, and the summation is over all SFTs in the data set
and over the frequency band 100--300~Hz.
We select the data set with the maximum value of the figure of merit.

At the time this search was conducted, SFTs were available from the beginning of S5 until 2007
April 19 UTC.  In this period, $\sim 9\%$ of the H1 science data and $\sim 13\%$ of the L1 science data
is excluded, either due to data quality vetoes or due to the segmentation of $h(t)$ during SFT
generation.
The 12-day data set selected for the search begins at 2007 March 20, 20:56:37 UTC and ends at
2007 April 1, 20:50:04 UTC.
It contains a total of 934 SFTs (445 from L1 and 489 from H1) and an
average of 9.7 days of data from each detector.

\subsection{Analysis Method}\label{sec:search:Fstat}

The data is searched using the $\F$-statistic, a coherent matched
filtering technique used to search for periodic gravitational waves using multiple detectors
\citep{JKS-1998,Cutler-Schutz-2005}.  
Matched filtering requires an accurate model, or template, of the signal.
The template models the response of
a detector to the two polarizations (``$+$'' and ``$\times$'') of the gravitational wave signal
emitted by a rotating neutron star.
In addition to the sky position, the barycentered gravitational wave
frequency, and its derivatives, the signal template has four parameters related
to amplitude and polarization:
the intrinsic strain $h_0$, initial phase constant $\phi_0$, inclination angle
$\iota$ of the star's rotation axis to the line of sight, and polarization
angle $\psi$. 
The $\F$-statistic
is the logarithm of the likelihood ratio analytically maximized over these unknown parameters.  The
value of the $\F$-statistic is usually quoted as $2\F$.

In transforming a signal from the reference frame of the star to that of the detector, the
intrinsic frequency of the source is modulated by Doppler effects due to the sidereal and orbital
motion of Earth with respect to the source.  The transformation requires the right ascension
$\alpha$ and declination $\delta$ of the star, which for Cas~A are known to high precision:
$\alpha = 23^\textrm{h} 23^\textrm{m} 27^\textrm{s}.943 \pm 0^\textrm{s}.05$, and $\delta =
58^{\circ} 48' 42''.51 \pm 0''.4$ \citep{Fesen-etal-limits-2006}.  The sky resolution of the
$\F$-statistic is, for a data set spanning 12 days, much coarser than the
measured uncertainties in $\alpha$ and $\delta$ \citep{Whitbeck-thesis-2006}, and so no search over sky position is required.  The
search is conducted over the remaining unknown function in the signal template: the instantaneous frequency of
the source as observed at the solar system barycenter, $f(t)$.  This is modeled as
\begin{equation}
f(t) \approx f + \fdot (t-t_0) + \frac{1}{2} \fdotdot (t-t_0)^2 \,,
\end{equation}
where the initial frequency $f$, first spin-down $\fdot$, and second spin-down $\fdotdot$ (all
evaluated at the start time of the data set, $t = t_0$) constitute the search parameters.
In previous searches for periodic
gravitational waves, it has not been necessary to include a second spin-down, but one is required for
this search due to the young age of Cas~A \citep{CasA-methods-2008}.

The signal template does not allow for the possibility that Cas~A glitched during the
12 days spanned by the data.
On the other hand, even the most frequently glitching pulsar does so only a few times per
year, and glitch population statistics do not clearly indicate that the youth
of Cas~A is sure to mean more frequent glitches \citep{YuanEtAl2010}.
The worst case would be a glitch at the midpoint of the 12 days span, since in
other cases the template would pick up the longer of the pre- and post-glitch
coherent stretches.
Amplitude signal-to-noise accumulates as the square root of observation time,
so the worst loss would be a factor of~$\sim 2$.

The $\F$-statistic implicitly assumes a uniform prior on
the polarization angle $\psi$, and
therefore the standard data analysis and upper limit procedures remain valid
for gravitational waves signals from $r$-modes, even
though the signal template used in this search assumes a mass quadrupole \citep{rpol}.

The Cas~A search uses the \texttt{ComputeFStatistic\_v2}
implementation of the $\F$-statistic, which is available as part of
the LALSuite software package.\footnote{%
The version of the software used in the search is
tagged with the identifier \texttt{S5CasASearch}.
See \url{https://www.lsc-group.phys.uwm.edu/daswg/projects/lalsuite.html}.}
Values of the $\F$-statistic returned by \texttt{ComputeFStatistic\_v2}
have an uncertainty of up to 5\%, due to
practical computation issues and optimizations \citep[see][]{Prix-CFSv2-2010}.

\subsection{Parameter Space}\label{sec:search:param}

\begin{figure}
%\autoconvpsfrag[texpos=c,epspos=r]{f0}{\small $f \, / \, \textrm{Hz}$}
%\autoconvpsfrag[texpos=r,epspos=r]{f1}{\small $\fdot \, / \, 10^{-8} \, \textrm{Hz s}^{-1}$}
%\autoconvpsfrag[texpos=c,epspos=c,angle=90]{f2}{\small $\fdotdot \, / \, 10^{-17} \, \textrm{Hz s}^{-2}$}
%\autoconvoptions{margin=10pt}
%\autoconvpdfoptions{to=png}
\plotone{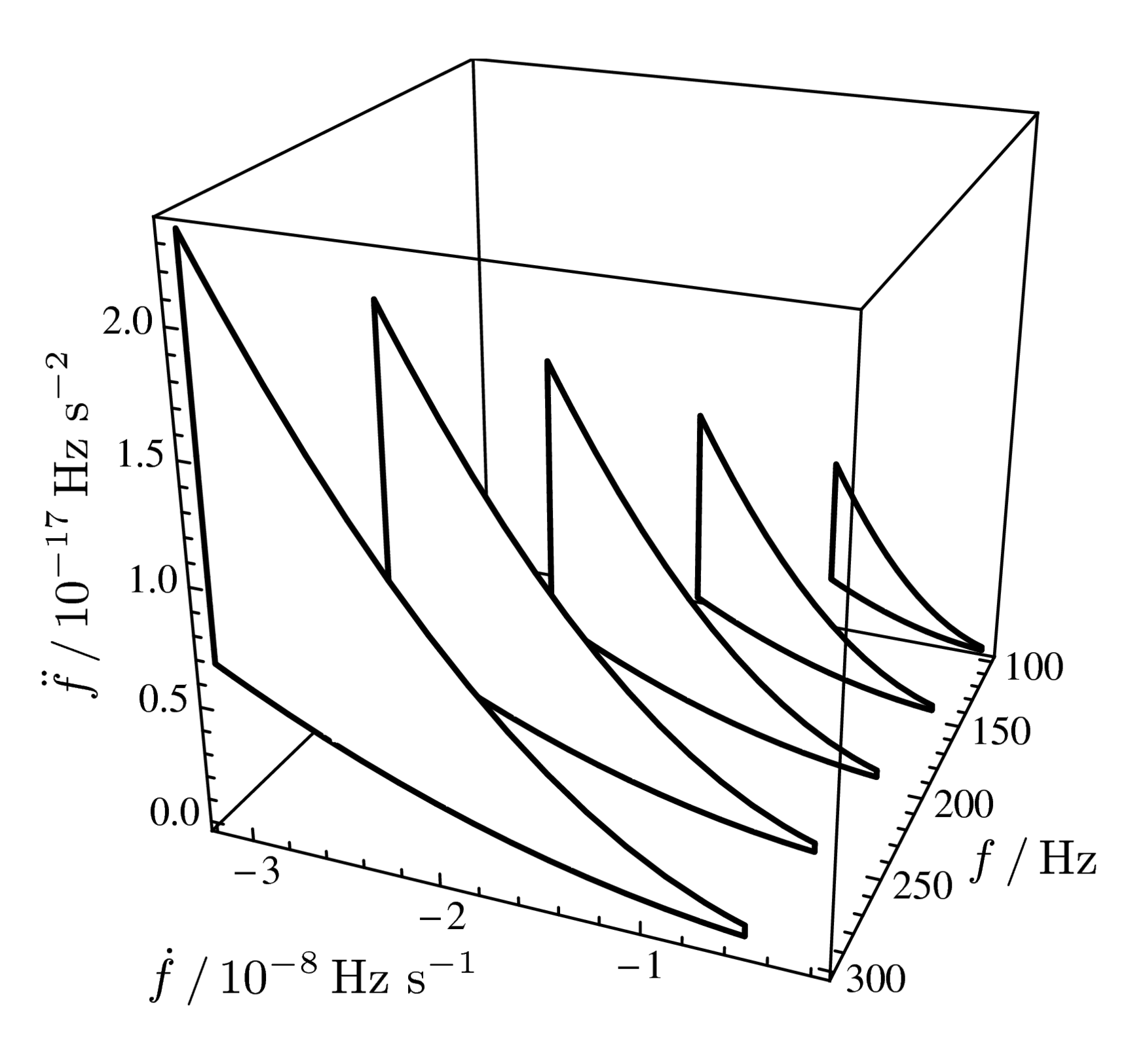}
\caption{
\label{fig:paramspace}
Visualization of the Cas~A search parameter space.  The black-outlined shapes are slices of the
$\fdot$--$\fdotdot$ parameter space at the following fixed values of $\f$: (back to front) 100, 150,
200, 250, and 300~Hz.  }
\end{figure}

The range of the gravitational wave frequency, $100~\textrm{Hz} \le f \le 300~\textrm{Hz}$, is
chosen based on the estimate in \citet{CasA-methods-2008} of the frequency band over which a search
of LIGO S5 data could beat the indirect limit on $h_0$ at reasonable computational cost.

The ranges of the spin-down parameters $\fdot$ and $\fdotdot$ are chosen to be \citep{CasA-methods-2008}
\begin{equation}
-\frac{f}{\langle \min n - 1 \rangle \tau}
\le \fdot \le
-\frac{f}{\langle \max n - 1 \rangle \tau}
\label{eqn:fdot}
\end{equation}
and
\begin{equation}
\frac{(\min n) \fdot^2}{f}
\le \fdotdot \le
\frac{(\max n) \fdot^2}{f}
\label{eqn:fdotdot}
\end{equation}
respectively, where the braking index $n$ is defined below. The age of Cas~A, $\tau$, is chosen to be 300~yr (as discussed in Section~\ref{sec:intro:gw}).
Note that the range
of $\fdot$ depends on $\f$, and the range of $\fdotdot$ depends on both $f$ and $\fdot$.
The
resulting shape of the three-dimensional parameter space of $f$, $\fdot$, and $\fdotdot$ is
depicted in Figure~\ref{fig:paramspace}.

The choices of $\fdot$ and $\fdotdot$ ranges are motivated by the desire to
cover a wide range of astrophysically motivated possibilities, expressed in
terms of the braking index.
In equation~\eqref{eqn:fdotdot}, we use the definition of the instantaneous braking
index $n = f \ddot{f} \dot{f}^{-2}$;
in equation~\eqref{eqn:fdot}, the angled brackets denote the average value of $n$
over the lifetime of Cas~A.
If the dominant emission mechanism driving the spin-down of Cas~A has changed
over its lifetime, these two quantities will be different.
Therefore, in the spirit of trying to cover the broadest imaginable parameter
space, we do not constrain the instantaneous and averaged values of $n$ to be
the same.
Instead, we search over values of $n$ (both instantaneous and averaged) between 2~and~7.
This range covers the possibilities that Cas~A is spinning down primarily due
to magnetic dipole radiation ($n = 3$) or to gravitational waves generated by
a mass quadrupole ($n = 5$) or by constant-$\alpha$ $r$-modes ($n = 7$).
This range
also covers the braking indices of nearly all known pulsars, which are typically between 2~and~3
\citep{Livingstone-etal-2006}.  The exception is the Vela pulsar, for which $n \approx 1.4$
\citep{Lyne-etal-1996}.
Extending $n$ to lower values would dramatically increase
the computational cost of the search.

\subsection{Computation}\label{sec:search:templ}

The search consists of computing the $\F$-statistic over a finite bank of templates, whose
parameters are given by points within the search parameter space.  The search uses a template bank
generation algorithm \citep{Wette-thesis-2009} which locates the parameter space points at the
vertices of a body-centered cubic lattice.
This minimizes the number of points per unit volume
required to cover the parameter space \citep[see, e.g.,][]{Conway-Sloane-1988,Prix-lattice-2007}.
The points are spaced using the $\F$-statistic parameter space metric
\citep{Whitbeck-thesis-2006,Prix-metric-2007} to ensure that the maximum
expected fractional loss in signal-to-noise ratio (known as the mismatch) will
never exceed 20\%.  Due to strong
correlations between the frequency and spin-down parameters, and the irregular shape of the
parameter space, it was necessary to place additional templates outside of the parameter space to
fully cover the parameter space boundaries, particularly for $\fdotdot$.  As a result, the number of
templates searched, $N \approx 7 \times 10^{12}$, is an order of magnitude larger than that
estimated in \citet{CasA-methods-2008}.
This resulted in a greater computational cost but does not greatly affect the
sensitivity of the search, which is only weakly dependent on the number of templates (see
Section~\ref{sec:search:2Fstar}).

The search is divided into $\sim 21500$ independent computational jobs by
partitioning the range of $f$ into small bands, each with approximately equal
numbers of templates, resulting in band widths of 1.8~mHz (at $f = 100$~Hz) to 20~mHz
(at $f = 300$~Hz).
The search
completed in $\sim 3.5$ days on $\sim 5000$ cores of the ATLAS computer cluster at the Max Planck
Institute for Gravitational Physics (Albert Einstein Institute) in Hanover, Germany.  From each
search job, only the $0.01\%$ of templates with the largest values of $2\F$ were recorded.

\subsection{Post-Processing}\label{sec:search:post}

\begin{table}
\caption{
\label{tab:veto-bands}
Frequency bands (column 1) identified during post-processing as containing spuriously large values
of the $\F$-statistic (the maximum of which are given in column 2), and a brief description of their
origin (column 3).  See the text for details.}
\begin{tabular}{lrl}
\hline
Frequency Band & $2\F_{\textrm{max}}$ & Origin \\
\hline
$108.860 \pm 0.018$ &  90 & Pulsar hardware injection no. 3 \\
$119.877 \pm 0.019$ &  72 & Sideband of 60 Hz harmonic \\
$128.000 \pm 0.017$ &  56 & 16 Hz harmonic \\
$139.225 \pm 0.058$ &  70 & L1-only line \\
$139.510 \pm 0.017$ &  72 & L1-only line \\
$144.751 \pm 0.056$ & 110 & L1-only line \\
$179.812 \pm 0.018$ &  51 & Sideband of 60 Hz harmonic \\
$185.630 \pm 0.055$ &  59 & L1-only line \\
$193.005 \pm 0.046$ &  66 & L1-only line \\
$193.391 \pm 0.018$ &  73 & Pulsar hardware injection no. 8 \\
$209.265 \pm 0.017$ &  54 & L1-only line \\
\hline
\end{tabular}
\end{table}

Data from the LIGO detectors are known to contain stationary or nearly stationary spectral lines originating from instrumental and
environmental noise.  The $\F$-statistic depends upon a robust estimator of the power spectral
density of the noise.
The implementation used in this search (see Section~\ref{sec:search:Fstat})
uses a spectral running median with a window size of 50 SFT bins, or 27.8 mHz
\citep{LSC-CW-Fstat-S2}.  Lines that are narrower than the median window size will remain in the data,
and may result in spuriously large values of the $\F$-statistic.

To identify frequency bands where the $\F$-statistic may have been contaminated in this manner, we search for
prominent narrow lines in power spectra of the searched H1 and L1 data.
This procedure identified eleven frequency bands, which
are listed in Table~\ref{tab:veto-bands}, along with the largest value of the $\F$-statistic found in each band.
Templates whose instantaneous frequency $f(t)$ at any time falls within any of these bands are
excluded from the remainder of the search pipeline.  Approximately $2 \times 10^{6}$ templates (a fraction 
$\sim 3 \times 10^{-6}$ of the total number of templates) are excluded in this manner.
Four of the bands contain a largest value of $2\F$ in the range 51--59, which would not be regarded as statistically significant gravitational wave candidates (see Section~\ref{sec:search:2Fstar}).

At certain times during S5, ten simulated periodic gravitational wave signals were injected into the
LIGO detectors at the hardware level by mechanically oscillating the detector mirrors
\citep{LSC-CW-PSH-S4}.  Four of the injections had frequencies within the search band.  The Cas~A
search data set was not selected with regard to times when the hardware injections were active.
As a
result, $\lesssim 1$~day of the searched data
($\sim 4.2\%$ of the H1 data, and $\sim 9.2\%$ of the L1 data)
contain the injections, and the effective intrinsic strains of the injections are reduced by a
factor of $\sim 10$.  After accounting for this reduction, two of the injections
have strains of $h_0 \lesssim 10^{-25}$, and are undetectable by this search.
The remaining two injections (designated nos.~3 and 8) have reduced
strains of $h_0 \sim 1.63 \times 10^{-24}$ and $1.59 \times 10^{-24}$ respectively.
These injections
are found by the Cas~A search at frequencies consistent with the parameters of the injected signals \citep{Wette-thesis-2009}.
Neither injection is at the sky position of Cas~A,
but are nevertheless detected due to
their strength, and the poor sky localization of short-duration (i.e. less than 1~day) signals
arising from
global correlations in the signal parameter space
\citep{Prix-Itoh-2005,Pletsch-2008}

The remaining nine non-injection bands are ruled out as gravitational wave candidates for
the following reasons.
Two contain harmonics of the 60 Hz power mains frequency, and one
contains a harmonic of the 16 Hz data acquisition buffering frequency \citep{LSC-stochastic-S1}.
The remaining six bands contain narrow instrumental lines which occur only in the L1 detector.
We would expect a convincing gravitational wave candidate to be seen in both
detectors.  Four of these six lines 
(at $\sim 139.2$~Hz, $144.7$~Hz, $185.6$~Hz, and $193.0$~Hz)
are definitively identified with environmental noise, by correlating the gravitational wave
data with data from environmental monitors (e.g.\ magnetometers, accelerometers and microphones).
The sources of the remaining two lines, at $\sim 139.5$~Hz and at $\sim 209.2$~Hz, are not
conclusively identified.  The value of the $\F$-statistic associated with the $\sim 209.2$~Hz line,
$2\F = 54$, is smaller than the largest value of $2\F$ expected from the search in the absence of
a gravitational wave signal (see next section), and is therefore not a candidate for a gravitational
wave signal from Cas~A.  The value of the $\F$-statistic associated with the $\sim 139.5$~Hz line,
$2\F = 72$, is also not statistically significant; as may be deduced from
Figure~\ref{fig:loudest-2F}, there is a $\sim 5\%$ probability that the search would return a higher
largest value of $2\F$ without a gravitational wave signal being present in the data.

\subsection{Significance of the Largest $2\F$}\label{sec:search:2Fstar}

The largest value of the $\F$-statistic returned by the search after post-processing is denoted
$2\Fstar$.
This is our most promising candidate for a gravitational wave signal from Cas~A.  The
probability density of $2\Fstar$, under the assumption that no gravitational wave signal
from Cas~A is present in the searched data, is given by
\begin{equation}
\label{eqn:dist-2Fstar}
p(2\Fstar) = N p(\chi^2_4; 2\Fstar) \left[
\int_{0}^{2\Fstar} \hspace{-1.5em} d(2\F) \, p(\chi^2_4; 2\F) \right]^{N-1} \,,
\end{equation}
where $N$ is the number of searched templates, and $p(\chi^2_4; 2\F)$ denotes the probability
density of a central $\chi^2$ distribution with 4 degrees of freedom
(i.e. the distribution of $2\F$ in the absence of any signal, and assuming Gaussian noise).
If the value of
$2\Fstar$ returned by the Cas~A search (see Section~\ref{sec:results}) is within the range of
probable values expected from $p(2\Fstar)$, it is not statistically significant.  If, on the other
hand, the value of $2\Fstar$ is extremely unlikely to be drawn from $p(2\Fstar)$, the candidate
signal is worthy of further investigation.

Equation~\eqref{eqn:dist-2Fstar} assumes that each of the $N$ values of $2\F$ are statistically
independent.
This is not strictly true, however, as we expect the $2\F$ values of neighbouring
templates to be correlated due to the close template spacing.  Therefore, in
equation~\eqref{eqn:dist-2Fstar}, $N$ should be substituted with the number of statistically
independent templates, $N_i < N$.  An empirical estimate of the number of statistically independent
templates found that $N_i \approx 0.88N$ for this search \citep{Wette-thesis-2009}.  A reduction in
$N$ shifts the distribution of $p(2\Fstar)$ towards lower values of $2\F$, and thus increases the
statistical significance of a candidate signal.  For a reduction of $N$ to $0.88N$, however, the
shift is negligible, and can be ignored; indeed, Figure~\ref{fig:loudest-2F} plots
equation~\eqref{eqn:dist-2Fstar} with $N_i = N$.  If, taking an extreme example, $N_i = 0.1N$, the
position of the maximum of $p(2\Fstar)$ would be shifted from $2\F \sim 66$ to $\sim 62$ (see
Figure~\ref{fig:loudest-2F}), but the significance of the $2\Fstar$ found by this search (see
Section~\ref{sec:results}) would not be greatly increased.
 Therefore, the uncertainty in the
number of statistically independent templates does not alter the significance of $2\Fstar$, nor the
conclusions reached in Section~\ref{sec:results}.  The determination the number of statistically
independent templates from first principles is an interesting area for further investigation.

\subsection{Upper Limits}\label{sec:search:ul}

If, after examining $2\Fstar$ (see Section~\ref{sec:search:2Fstar}),
we conclude that no gravitational signal has been detected,
we proceed to set 95\% confidence upper limits on the
intrinsic strain $h_0$, the ellipticity $\epsilon$, and the $r$-mode amplitude $\alpha$.  The upper
limits are determined using Monte Carlo injections, following the procedure described in
\citet{LSC-CW-Fstat-S2}.  The search frequency band is first partitioned into 400 sub-bands of width
0.5~Hz; an upper limit is set separately for each sub-band. The choice of 0.5~Hz is small enough
that the time-averaged noise floor of the detectors is approximately constant over each band,
and is large enough to keep the computational cost reasonable.
We denote the largest values of $2\F$ found by the full Cas~A
search in each sub-band by $2\Fstar_{\textrm{s-b}}$; they serve as the false alarm thresholds in each
band.  For each sub-band, a population of 5000--6000 periodic gravitational wave signals, with $h_0$
fixed and all other parameters randomly chosen, are each in turn injected into the search data set
and then searched for using the same analysis method used in the full Cas~A search. We record the
largest value of $2\F$ found by the searches for each of the injected signals. The fraction of
these $2\F$ values that are greater than $2\Fstar_{\textrm{s-b}}$ gives the confidence corresponding to the
fixed value of $h_0$.

To expedite the injection procedure, we use an analytic model of the distribution of the population
of injected signals to provide an educated guess at the value of $h_0$ required for 95\% confidence
\citep{Wette-thesis-2009}.  We then perform the injection procedure once, at that $h_0$, to check
that the recovered fraction really is 95\%. We find that the analytic model slightly overestimates
the required $h_0$, and so the 95\% confidence upper limits presented in Section~\ref{sec:results}
are conservative.

\section{Results}\label{sec:results}

\begin{figure}
%\autoconvpsfrag[texpos=c,epspos=c]{PSF2F}{$2\Fstar$}
%\autoconvpsfrag[texpos=c,epspos=c]{PSFp2F}{$p(2\Fstar)$}
%\autoconvpsfrag[texpos=c,epspos=c]{PSF2Fstar}{$2\Fstar \approx 65$}
\plotone{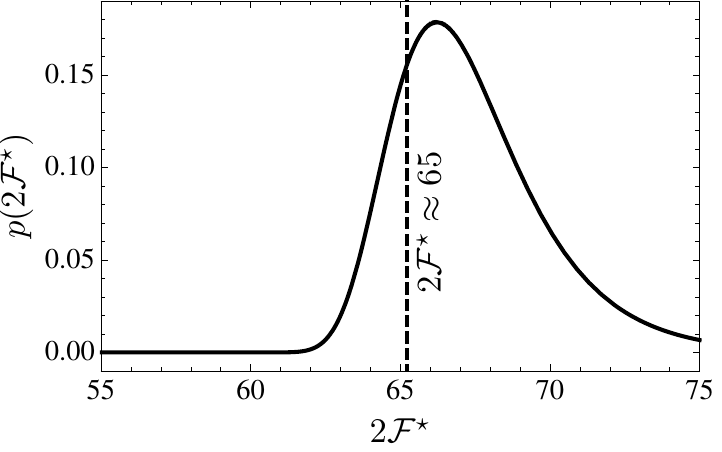}
\caption{
\label{fig:loudest-2F}
The solid line is the probability density of the largest value of the $\F$-statistic,
$2\Fstar$, under the assumption that no gravitational wave signal from Cas~A is present in the
searched data [see equation~\eqref{eqn:dist-2Fstar}].  The dashed line denotes the value of
$2\Fstar$ returned by the Cas~A search.}
\end{figure}

After post-processing, including the exclusion of the 11 frequency bands discussed in section~\ref{sec:search:post},
the largest remaining value of the $\F$-statistic returned by the search is
$2\Fstar \approx 65$. Figure~\ref{fig:loudest-2F} compares $2\Fstar$ to its expected theoretical
distribution, under the assumption that the searched data contains no gravitational wave signal from
Cas~A (see Section~\ref{sec:search:2Fstar}).  It is clear that $2\Fstar$ is consistent with this
distribution, and is therefore not statistically significant.  We therefore conclude that the
searched data does not contain any plausible gravitational wave signal from Cas~A.
This conclusion is not significantly influenced by
the uncertainty in the number of statistically independent templates (see
Section~\ref{sec:search:2Fstar}).

\begin{figure}
%\autoconvpsfrag[texpos=c,epspos=c]{PSFf}{Gravitational wave frequency / Hz}
%\autoconvpsfrag[texpos=c,epspos=c]{PSFTms}{Rotation period / ms}
%\autoconvpsfrag[texpos=c,epspos=c]{PSFh0m24}{Intrinsic strain $h_0$ / $10^{-24}$}
\plotone{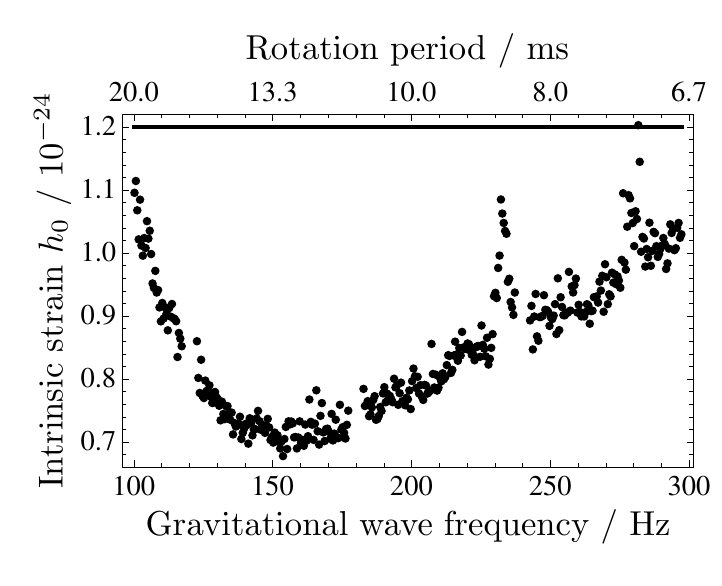}
\caption{
\label{fig:ul-h0}
Upper limits at 95\% confidence (dots) on the intrinsic strain $h_0$ of gravitational waves from
Cas~A, and the indirect limit (line). The gravitational wave frequency is assumed to be twice the
rotation frequency.
Systematic uncertainties are not included; see Section~\ref{sec:results} for
discussion.
}
\end{figure}

\begin{figure}
%\autoconvpsfrag[texpos=c,epspos=c]{PSFf}{Gravitational wave frequency / Hz}
%\autoconvpsfrag[texpos=c,epspos=c]{PSFTms}{Rotation period / ms}
%\autoconvpsfrag[texpos=c,epspos=c]{PSFepsm4}{Ellipticity $\epsilon$ / $10^{-4}$}
\plotone{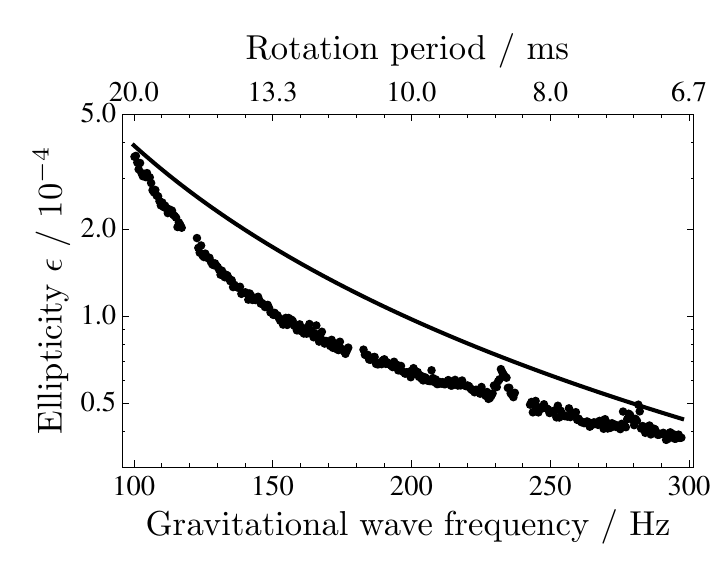}
\caption{
\label{fig:ul-eps}
Upper limits at 95\% confidence (dots) on the equatorial ellipticity $\epsilon$ of Cas~A, and the
indirect limit (line). The gravitational wave frequency is assumed to be twice the rotation
frequency.
Systematic uncertainties are not included; see Section~\ref{sec:results} for
discussion.
}
\end{figure}

\begin{figure}
%\autoconvpsfrag[texpos=c,epspos=c]{PSFf}{Gravitational wave frequency / Hz}
%\autoconvpsfrag[texpos=c,epspos=c]{PSFTms}{Rotation period / ms}
%\autoconvpsfrag[texpos=c,epspos=c]{PSFalpha}{Amplitude of $r$-mode $\alpha$}
\plotone{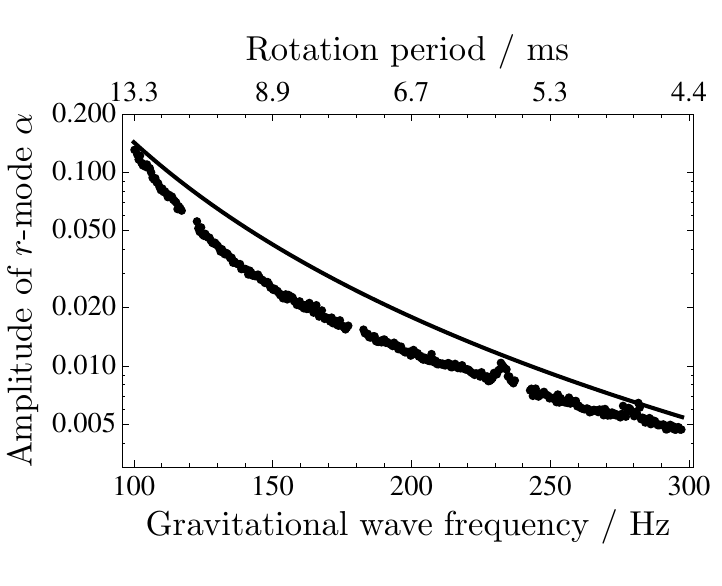}
\caption{
\label{fig:ul-alpha}
Upper limits at 95\% confidence (dots) on the amplitude $\alpha$ of $r$-mode oscillations of Cas~A,
and the indirect limit (line). The gravitational wave frequency is assumed to be 4/3 times the
rotation frequency.
Systematic uncertainties are not included; see Section~\ref{sec:results} for
discussion.
}
\end{figure}

Upper limits at 95\% statistical confidence on $h_0$, $\epsilon$, and $\alpha$ are plotted, alongside their
respective indirect limits, in Figures~\ref{fig:ul-h0}, \ref{fig:ul-eps}, and~\ref{fig:ul-alpha}
respectively.  
Systematic uncertainties in the direct upper limits on $h_0$ are of order 10--15\%
(see Sections~\ref{sec:search:data} and~\ref{sec:search:Fstat}).
Systematic uncertainties in all three indirect limits, and in the direct upper limits
on $\epsilon$ and $\alpha$ inferred from $h_0$, are roughly a factor 2--3 for
$\alpha$ and 2 for the others (see Section~\ref{sec:intro:gw}).
As expected, the upper limits beat the indirect limits over the gravitational wave
frequencies 100--300~Hz. The equivalent ranges of rotation periods are 6.7--20~ms for the upper
limits on $h_0$ and $\epsilon$ (which assume the gravitational wave frequency to be twice the
rotation frequency), and 4.4--13~ms for the upper limits on $\alpha$ (which assume a 4/3 ratio of
gravitational wave frequency to rotation frequency).
Upper limits within 2~Hz of harmonics of the 60~Hz power mains frequency are severely degraded due to noise, and are excluded from the figures.
Between 230--240~Hz, and around 280~Hz, the upper limits are degraded by disturbances in the broadband noise of the H1 detector.

\section{Discussion}\label{sec:discuss}

The tightest upper limit on $h_0$ (Figure~\ref{fig:ul-h0}) is $\sim 7 \times 10^{-25}$ at $\sim
150$~Hz, in the region where the LIGO detectors are at their most
sensitive.  The search improved slightly upon the expected upper limits on $h_0$ estimated by
\citet{CasA-methods-2008}, due to slightly better detector sensitivity and duty cycle in the
selected data set.
Therefore, the search could have beaten the indirect limits at frequencies slightly outside
of the 100--300~Hz band, although the computational cost increases rapidly at higher frequencies,
and the noise floor of the detectors rises steeply at lower frequencies.

The upper limits on $\epsilon$ (Figure~\ref{fig:ul-eps}) range from $\sim 4 \times 10^{-4}$ at
100~Hz (20~ms rotation period) to $\sim 4 \times 10^{-5}$ at 300~Hz (6.7~ms),
assuming the canonical parameters of Eq.~\eqref{eqn:ul-eps}.
The upper limits are higher than the maximum $\epsilon$ of a few times $10^{-6}$
predicted for normal neutron stars, even with recent results indicating a high
breaking strain of the crust \citep{Horowitz-Kadau-2009}.
Ellipticities of a few times $10^{-4}$ are within the range of predictions
\citep{Owen-exotic-2005, Lin-2007, Haskell-etal-2007, Knippel-Sedrakian-2009}
for various forms of crystalline quark matter \citep{Xu-2003,
Mannarelli-etal-2007}.
Robust hybrid models \citep{Glendenning-1992} could sustain ellipticities up to
about $1\times10^{-4}$ \citep[scaled from][]{Owen-exotic-2005} if the breaking
strain of \citet{Horowitz-Kadau-2009} is valid for the mixed phase of matter.
Ellipticities comparable to our upper
limits could also be sustained by internal magnetic fields of order $10^{16}$~G, depending on the
field configuration, equation of state, and superconductivity of the star
\citep{Cutler-2002,Haskell-etal-2008,Akgun-Wasserman-2007,Colaiuda-etal-2007}.

It is important to realise that upper limits on $\epsilon$ cannot be used to
constrain properties of QCD or the composition of the neutron star, which may
simply have an ellipticity much lower than the theoretical maximum.
The upper limits on $\epsilon$ do, however, constrain the internal magnetic
field to be less than of order $10^{16}$~G, if Cas~A is spinning fast enough
to radiate gravitational waves in the searched frequency band.

The upper limits on $\alpha$ (Figure~\ref{fig:ul-alpha}) range from $\sim 0.14$
at 100~Hz (13~ms rotation period) to $\sim 0.005$ at 300~Hz (4.4~ms), assuming
the canonical parameters of Eq.~\eqref{eqn:ul-alpha}.
If the $r$-mode amplitude varies with time, our limits on $\alpha$ are rms
values over the observing time and the indirect limits are rms values over the
lifetime of the star.
Our upper limits on $\alpha$ are within the range of runaway low-viscosity
scenarios at all frequencies, and on the high end of the frequency band they
are comparable to the finite-viscosity parametric instability thresholds which
tend to serve as attractors for the evolution \citep{Bondarescu-etal-2009}.

In several years the advanced LIGO and Virgo interferometers are expected to be in operation, with
sensitivities an order of magnitude better than data searched here and extending
to lower frequencies.  Extrapolating from these
results, a similar search on data from advanced interferometers would be expected to be sensitive to
ellipticities of a few times $10^{-6}$, which are achievable by neutron stars without exotic matter
or by stars with internal magnetic fields less than $\sim 10^{14}$~G, or $r$-mode amplitudes a few times $10^{-4}$.  More sophisticated data
analysis methods, such as hierarchical methods, would further increase the sensitivity.  There are
also more young non-pulsing neutron stars and other astrophysically interesting objects that could
be targeted by a search of this type.

\acknowledgements

The authors gratefully acknowledge the support of the United States
National Science Foundation for the construction and operation of the
LIGO Laboratory and the Science and Technology Facilities Council of the
United Kingdom, the Max-Planck-Society, and the State of
Niedersachsen/Germany for support of the construction and operation of
the GEO600 detector. The authors also gratefully acknowledge the support
of the research by these agencies and by the Australian Research Council,
the Council of Scientific and Industrial Research of India, the Istituto
Nazionale di Fisica Nucleare of Italy, the Spanish Ministerio de
Educaci\'on y Ciencia, the Conselleria d'Economia, Hisenda i Innovaci\'o of
the Govern de les Illes Balears, the Royal Society, the Scottish Funding 
Council, the Scottish Universities Physics Alliance, The National Aeronautics 
and Space Administration, the Carnegie Trust, the Leverhulme Trust, the David
and Lucile Packard Foundation, the Research Corporation, and the Alfred
P. Sloan Foundation.

This paper has been designated LIGO Document No.~\LIGOdcc.

\bibliography{apj-jour,CasAResultsPaper}

\end{document}